\documentclass[11pt]{article}
\usepackage{moriond,epsfig}

\def\be{\begin{equation}}
\def\ee{\end{equation}}
\def\bea{\begin{eqnarray}}
\def\eea{\end{eqnarray}}

\begin{document}
\vspace*{4cm}
\title{DGLAP IMPROVED SATURATION MODEL WITH HEAVY FLAVOURS}

\author{S.~SAPETA}

\address{M. Smoluchowski Institute of Physics, Jagellonian University \\
         Reymonta 4, 30-059 Krak\'ow, Poland}

\maketitle\abstracts{
The DGLAP improved saturation model of Bartels, Golec
and Kowalski is supplemented by the contribution from the heavy quarks: charm and beauty.  
It is shown to give good description of both the total proton structure function $F_2$ and, as a prediction, the heavy quarks contribution $F^{c\bar{c}}_2$ and $F^{b\bar{b}}_2$. The reasonable agreement for the longitudinal and diffractive structure functions is also found.
}


Among a number of the dipole models of deep inelastic scattering the one proposed by Golec-Biernat and W\"usthoff \cite{Golec-Biernat:1998js}$^,$~\cite{Golec-Biernat:1999qd} turned out to be especially popular. This is because it seems to grasp the essential elements of parton saturation being at the same time relatively simple.  It gave a successful fit to low $x$ HERA data for the structure function $F_2$. However, with the advent of new more precise data the model required some improvement. This has been done by Bartels, Golec-Biernat and Kowalski \cite{Bartels:2002cj} (BGK model) by taking into account DGLAP evolution of the gluon density in the proton, which affects the behavior of the dipole cross section for small dipole sizes $r$.

The photon--proton interaction at low $x$ my be divided into two stages. In the first stage the transversely or longitudinally polarized virtual photon  splits into a quark-anti-quark pair (a color dipole). This splitting is described by the photon wave function $\Psi(\vec{r},z,Q^2,m^2_f)$, where $z$ denotes the fraction of the photon momentum carried by quark in the dipole. In the second stage the dipole interacts with the proton with the dipole cross section $\hat\sigma(x,\vec{r})$ which is usually modeled using ideas from high density QCD \cite{kovchegov}. 
In such description $F_2$ is a sum of the contributions $F^f_2$ coming from different quark flavours in the color dipole

\begin{equation}
\label{eq:f2sum}
F_2 (x, Q^2)= \sum_f F^f_2 (x, Q^2)
\end{equation}
and
\begin{equation}
F^f_2(x, Q^2)
= \frac{Q^2}{4\pi^2\, \alpha_{em}}\,
  \sum_{P} \int \!d\,^2\vec{r}\! \int_0^1 \!dz\;
  \vert \Psi_{P}^f\,(\vec{r},z,Q^2,m^2_f) \vert ^2 \: 
  \hat{\sigma}\,(x,\vec{r}) 
\end{equation}
where $P$ is the virtual photon polarization.

In the BGK model the dipole cross section has the form
\begin{equation}
\hat{\sigma}(x,r)  = \sigma_0 \left\{1-\exp\left(-
                     \frac{ \pi^2}{3\, \sigma_0}\, r^2\, \alpha_s(\mu^2)\, 
                           xg(x,\mu^2) 
                     \right)\right\}
\end{equation}
with the scale $\mu^2$ assumed to run with $r^2$ as $\mu^2 = C/r^2 + {\mu^2_0}$
and the gluon density $xg(x,\mu^2)$ which evolves with $\mu^2$ according to the DGLAP equations with the initial condition
\begin{equation}
\label{eq:gluon_int}
xg(x,Q^2_0) =  A_g \, x^{- \lambda_g}(1-x)^{5.6}
\qquad {\rm at} \quad Q^2_0 = 1\ {\rm GeV}^2
\end{equation}
The model has five parameters $\sigma_0$, $C$, $\mu^2_0$, $A_g$ and $\lambda_g$, which are fitted to the $F_2$ data.  In the paper of Bartels Golec-Biernat and Kowalski~\cite{Bartels:2002cj} only the three light quarks were considered in the sum (\ref{eq:f2sum}). Thus the main goal of the work presented here is to consider the charm and beauty contributions within the BGK model. Notice that after the model parameters have been fitted to the $F_2$ data, the contributions $F^{c\bar{c}}_2$ and $F^{b\bar{b}}_2$ will come out as a prediction.

One should emphasize that the heavy quarks production in deep inelastic scattering at low~$x$ could by no means be neglected. According to H1 \cite{Adloff:plb528}$^,$ \cite{Aktas:epjc40}$^,$ \cite{H1-prel-heavy} and ZEUS \cite{Chekanov:prd69} the contribution of the charm structure function to the total structure function varies from $10\%$ to $30\%$ while for the beauty structure function it reaches up to $3\%$.

We performed fits with heavy quarks using the most recent H1 \cite{Adloff:epjc21} and ZEUS \cite{Chekanov:epjc21}$^,$ \cite{Breitweg:plb487} inclusive DIS data with $x_B \leq 0.01$ and $0.1 \leq  Q^2 \leq 4000\, {\rm GeV}^2$. We took the light quarks massless and the typical values of masses for heavy quarks: $m_{\rm charm}     =  1.3\ {\rm GeV}$,  $m_{\rm beauty}  =  5.0\ {\rm GeV}$ and modified the Bjorken variable to be $x = x_B(1+4m^2_f/Q^2)$, more details are given elsewhere \cite{Golec-Sapeta}. 
The fit results are presented in Table \ref{fit_results}. In the first row we recall the parameters obtained in the BGK fit \cite{Bartels:2002cj} with light quarks only. Our fits with heavy quarks  are presented in the next two lines.
As we see from the value of $\chi^2/{\rm ndf}$ the overall fit quality is still good. However, the values of parameters concerning the gluon differ significantly from the light quarks fit. 
In particular, $\lambda_g$ is positive which means that the initial gluon distribution~(\ref{eq:gluon_int}) grows with decreasing $x$, which is opposite to the case with light quarks only when this distribution is valence--like.

\begin{figure}[p]
\begin{center}
\psfig{figure=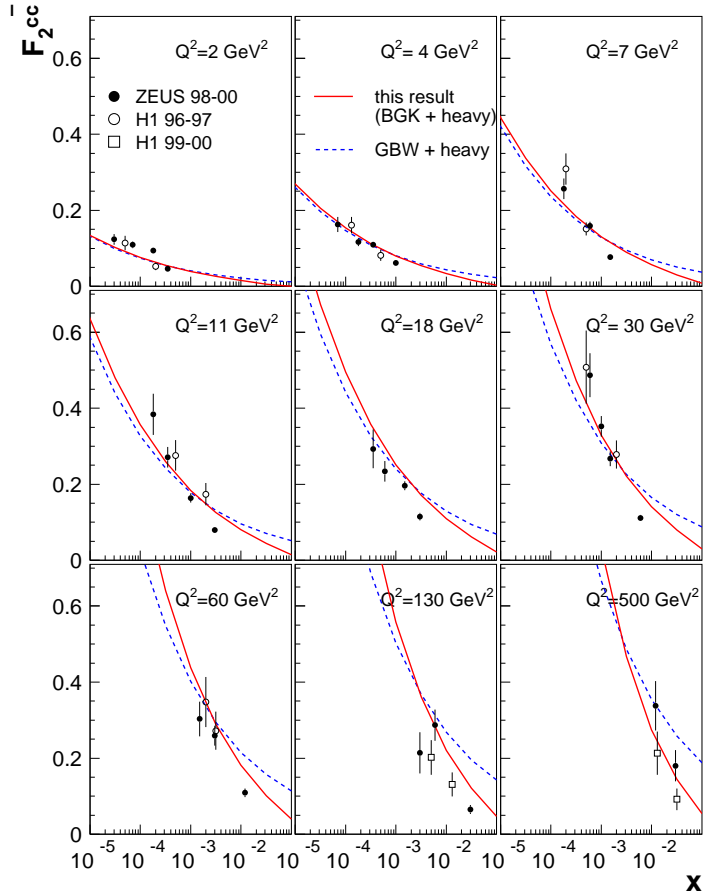,height=5.1in}
\caption{Predictions for the charm structure function $F^{c\bar{c}}_2$ in the DGLAP improved saturation model (solid lines). The predictions in the old GBW model without DGLAP evolution are shown for reference (dashed lines).} 
\label{fig:f2charm}
\end{center}
\end{figure}

\begin{figure}[p]
\begin{center}
\psfig{figure=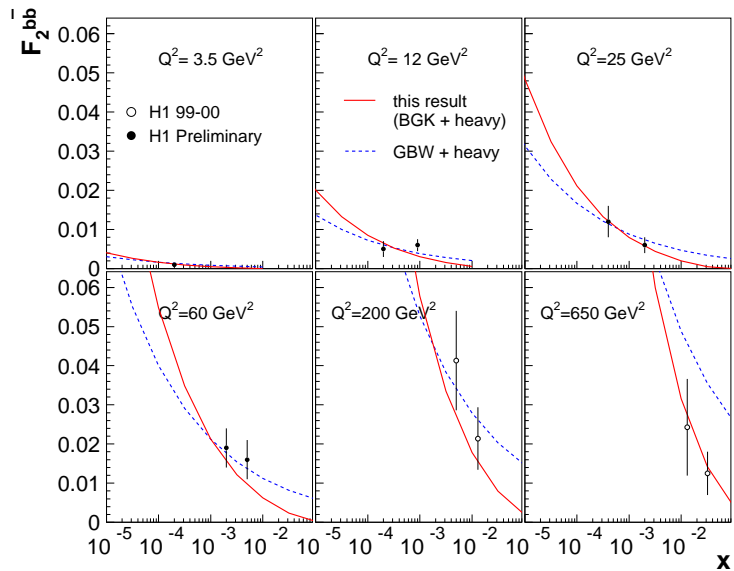,height=3.3in}
\caption{The same as in Fig.~\ref{fig:f2charm} but for the beauty structure function $F^{b\bar{b}}_2$.}
\label{fig:f2beauty}
\end{center}
\end{figure}

\begin{table}[t]
\caption{Parameters of the DGLAP improved saturation model with heavy quarks obtained from the fit to $F_2$. The results from the BGK fit with light quarks only are shown for reference in the first row.}
\vspace{0.3cm}
\begin{center}
\begin{tabular}{|c||c|c|c|c|c||c|} \hline 
& $\sigma_0\,$[mb] & $A_g$ & $\lambda_g$ & $C$ & $\mu^2_0$ & $\chi^2/ndf$
\\ \hline  \hline 
light \cite{Bartels:2002cj}    & 23.8 & 13.71 & - 0.41 & 11.10 & 0.52 & 0.97
\\ \hline \hline
light + c                      & 22.4 & 1.78  & 0.0803 & 0.679 & 2.11 & 1.15
\\ \hline
light + c + b                  & 22.5 & 1.77  & 0.0793 & 0.761 & 2.08 & 1.25
\\ \hline
\end{tabular}
\end{center}
\vspace{-0.2cm}
\label{fit_results}
\end{table}
With the obtained parameters we are able to give predictions for the charm and beauty structure functions $F^{c\bar{c}}_2$ and $F^{b\bar{b}}_2$. The results are plotted in Figs. \ref{fig:f2charm} and \ref{fig:f2beauty} (solid lines). As a reference we show also the old GBW results without DGLAP modifications (dashed lines). We observe good agreement both in the normalization and in the slope for all $Q^2$ bins. The latter is improved by the DGLAP evolution, which is especially important at high $Q^2$.

The model provides also the predictions for the longitudinal structure function $F_L$ and the diffractive structure function $F^{D(3)}_2$. The obtained values of $F_L$ are in agreement with H1 estimations~\cite{Adloff:epjc21}$^,$ \cite{Adloff:2003uh}$^,$ \cite{H1-prel-fl}. The comparison of this predictions with direct measurements, which are planned by H1 and ZEUS collaborations, will be particularly interesting. For the case of $F^{D(3)}_2$ we still obtain reasonable agreement with the ZEUS data \cite{Chekanov:2005vv}.  

In summary, we have shown that the DGLAP improved saturation model with heavy quarks gives successful description of $F_2$ at low $x$ and provides predictions for the charm and beauty structure functions $F^{c\bar{c}}_2$ and $F^{b\bar{b}}_2$, which nicely agree with the data. For more details we refer to the paper \cite{Golec-Sapeta}.


\section*{Acknowledgments}
This work was done in collaboration with Krzysztof Golec-Biernat. The research has been supported by the grant of Polish Ministry of Science and Information Society Technologies: No.~1~P03B~02828.

\end{document}